\documentstyle[12pt]{article}
\def\gev{{\rm \, Ge\kern-0.125em V}}
\begin{document}
\begin{titlepage}
\pagestyle{empty}
\baselineskip=21pt
\rightline{McGill 95--41}
\rightline{July 1995}
\vskip .2in
\begin{center}
{\large{\bf A No-Go Theorem in String Cosmology}}

\end{center}
\vskip .5in
\begin{center}
Nemanja Kaloper

{\it Department of Physics, McGill University}

{\it Montr\'eal, Qu\'ebec, Canada H3A 2T8}

{\it email: kaloper@hep.physics.mcgill.ca}

\vskip1in

\end{center}
\centerline{ {\bf Abstract} }
\baselineskip=18pt
A no-go theorem pertaining to the graceful exit problem in Pre-Big-Bang
inflation is reviewed. It is shown that dilaton self-interactions and
string fluid sources fail to facilitate branch changing necessary to
avoid singularities. A comment on the failure of the higher genus
corrections to induce graceful exit is also included.

\vskip1in
\begin{center}
Contributed talk at the $6^{th}$ Canadian Conference on
General Relativity and Relativistic Astrophysics,
University of New Brunswick, Fredericton, N.B, Canada May 24-28, 1995.
\end{center}
\end{titlepage}
\baselineskip=18pt
{\newcommand{\la}{\mbox{\raisebox{-.6ex}{~$\stackrel{<}{\sim}$~}}}}
{\newcommand{\ga}{\mbox{\raisebox{-.6ex}{~$\stackrel{>}{\sim}$~}}}}
\def\beq{\begin{equation}}
\def\eeq{\end{equation}}
Inflationary paradigm is one of the cornerstones of modern
cosmology \cite{infl}. It states that there must have occurred a stage
of rapid expansion
of the Universe, in order to account for such large scale properties as
homogeneity and isotropy, flatness, absence of topological relics etc.
A simple argument illustrating the need for inflation is the so called
horizon problem. Namely, we cannot reconcile the observed high degree of
correlation
in the background radiation (with the angular fluctuations in black-body
temperature of the order $\delta T/T \propto 10^{-6}$) with the evolution
of the Universe dictated by Einstein relativity and dominated by fluid
sources. The Universe must have evolved out of roughly $10^5$ causally
disconnected domains at the time of decoupling of radiation from matter
in order to achieve its present size.
As the radiation did not interact from that
time on, it is hard to see how the present correlation is attained.
Inflationary scenario saves the day by postulating an era
of rapid expansion prior to decoupling, producing a huge causally correlated
region from which our Universe evolved.

We still need to construct
a fully self-consistent dynamics of inflation,
in agreement with observations and free of internal inconsistencies.
A possibility may be found in string theory \cite{gsw}. Up to date,
it is the only candidate for a unified theory of interactions.
If we accept it for its promise, it is important to see if the
inflationary scenario, dictated by observations, can be incorporated
naturally in string cosmology.
This in fact could be viewed as a test of string theory.
It turns out that this is not an easy task. Typically there
appear new difficulties, such as the
rolling scalar fields and induced variation of particle masses and
couplings \cite{CLO}. In response, an alternative
approach towards string-driven inflation, dubbed the Pre-Big-Bang
inflation, has been suggested \cite{gv,bv}. It strives to induce inflation
deriving from genuinely stringy mechanisms, relying precisely on the
rolling dilaton.
The scenario is defined in the string world-sheet frame, where there exist two
branches of solutions related by string scale factor duality. They
are characterized by
superexponential, pole-driven inflation for $t<0$ and a milder,
power-law expansion
for $t>0$. If the jump (branch change) at  (or near) $t =0$ could be made
from the superinflationary phase
to the power-law one, the dreaded singularity may be avoided.
However, the scenario does not work in its simplest form \cite{bv,kmo}.
The desired branch
changes cannot be catalyzed by dilaton self-interactions and/or string
fluid sources.
Here we will review this recent result, and also will comment on the
failure of
the higher genus corrections introduced by Damour and Polyakov \cite{dp}
to induce such
branch changes \cite{kmo}. This review is based on the joint work of
R. Madden, K.A. Olive
and the author \cite{kmo}, where a more detailed account can be found.

We start with the simplest case, defined by the action
\beq
S = \int d^4x \sqrt{g} e^{-2 \phi} \Bigl\{\frac{R}{2}
 + 2 \partial_{\mu} \phi
\partial^{\mu} \phi - \Lambda(\phi)\Bigr\}
\label{sact}
\eeq
Here $\phi$ is the dilaton and $\Lambda(\phi)$ is the dilaton
potential. We have ignored the contribution of the
axion since during expansion an axion dominated universe can be
expected to quickly
evolve to one dominated by the dilaton.
Assuming a spatially flat Robertson-Walker geometry,
\beq
ds^2= -dt^2+a(t)^2 d{\vec x}^2
\eeq
we can write the two independent equations of motion in a relatively simple
first order form:
\begin{eqnarray}
\dot \phi &=& (3 h \pm \sqrt{3 h^2+2 \Lambda(\phi)})/2 \label{bv1} \\
\noalign{\medskip}
\dot h   &=& \pm h \sqrt{3 h^2+2 \Lambda(\phi)}-\Lambda'(\phi)/2
\label{bv2}
\end{eqnarray}
with the $\pm$ sign chosen for both equations simultaneously.
These equations are easily solved when $\Lambda=0$
resulting in four different solutions, two for each of the
two branches corresponding to the choice of
(+) and ($-$) sign. The (+) branch is defined in the
domain $t<0$ and the ($-$) branch in $t>0$.
The solutions are \cite{bv}
\begin{eqnarray}
&&a=a_0 |t|^{\mp \frac{1}{\sqrt{3}}}, ~~~ h = \mp \frac{1}{\sqrt{3} t},
 ~~~ \phi = \phi_0 + \frac{\mp \sqrt{3} - 1}{2} \ln |t|
{}~~~~ {\rm for} ~t<0 ~((+) ~{\rm branch}) \nonumber \\
&&a=a_0 t^{\pm \frac{1}{\sqrt{3}}}, ~~~~~ h = \pm \frac{1}{\sqrt{3} t},
 ~~~ \phi = \phi_0 + \frac{\pm \sqrt{3} - 1}{2} \ln t
{}~~~~~ {\rm for} ~t>0 ~((-) ~{\rm branch}) ~~~~~~
\label{bvsols}
\end{eqnarray}
We will focus here on the expanding solutions.
The expanding solution for $t<0$
begins in the weak coupling
regime ($\phi$ large and negative) and evolves
toward the strong coupling region
($\phi$ positive), compatible with the form of the action which
represents a weak-coupling truncation of the full effective
action of string theory. The expanding solution for $t>0$
can be matched to a standard post-inflationary Robertson-Walker
cosmology, by turning on dilaton self-interactions.
We repeat here that for $t<0$ we have a pole-driven
expansion, reaching the singularity at $t=0$,
and for $t>0$ we get a power-law
expanding universe emerging from the singularity.
If the singularity at the pole $t=0$ were removed and the two
branches joined smoothly,
the resulting solution would represent a completely nonsingular cosmology.
In the region of the order of Planck scale the superinflating branch
would metamorphose to a cooling,
expanding universe which can be joined onto our own
as $t \rightarrow \infty$ \cite{bv}. We note that the evolution
should eventually
lead to the decoupling of the dilaton by trapping it in a fixed
point (potential well)
in order to match the solution to a late time cosmology.
The high curvature region around $t=0$ would
resemble the Big Bang and therefore, in addition to possibly solving the
problems usually assigned to inflation, it would also give an elegant
resolution to the question of initial singularity.

The mechanism of branch changing should be responsible for
graceful exit. We now turn to it. It is obvious from the equations of
motion that the two branches
can never connect smoothly in the regions where the
potential is positive (cf. eqs. (\ref{bv1})
and (\ref{bv2})) since we must match derivatives at the point of
contact. This requires that
the potential is negative in a certain region.
If we represent the dynamics by the phase space portrait in the phase plane
$(\phi, h)$, the regions where branch changes can occur are closed
curves symmetric around the $\phi$-axis, given by $3h^2 +2\Lambda = 0$.
They were conveniently named the ``eggs" because of their
concave shape in the regions containing a single negative minimum of
$\Lambda$.
Unfortunately, in our case although eggs lead to branch changes,
there is no graceful exit since branch changes always occur in
pairs. To see it, we define the egg function:
$e=\sqrt{3h^2+2 \Lambda(\phi)}$, which satisfies
$\dot e=\pm (2 h \dot \phi-\dot h)$.
By dividing both sides
by $\dot \phi$ and integrating the result over $\phi$ along a trajectory,
we obtain:
\begin{eqnarray}
\pm(e(t_1)-e(t_0))+h(t_1)-h(t_0)=2 \int_{\phi(t_0)}^{\phi(t_1)} h d\phi
\label{diffint}
\end{eqnarray}
The integral here is to be understood as a line integral along the path
of the system between $\phi(t_0)$ and $\phi(t_1)$. In fact this is the
equation of phase trajectory in integral form. The positivity of the integral
in relevant cases is the input needed to obtain the no-go theorem.

This comes about as follows. From looking at the flows in the phase space,
we see that any trajectory hitting the top of the
egg must have come from the left, and any trajectory hitting below
the egg must have come from the right. A case when a trajectory flows
around the egg without hitting it for ``half" a cycle is also
consistent with the above remark.
Then we can show that 1) no ($-$) branch trajectory,
originating from anywhere on the upper side of the
egg can escape over the right end of the egg but must rehit it, and
2) no (+) trajectory coming from the right and flowing below the
egg hits the egg below, or on,
the $\phi$-axis.
{}From this, we conclude that
any (+) branch entering an egg region from the left must go over the top
of the egg, possibly experiencing several branch changes,
and must exit the region of the egg while still on the (+) branch.
No (+) branch entering an egg region from the right
can hit below, and it must remain (+) while flowing
under the egg. Thus any (+) trajectory entering the egg region
cannot leave on a ($-$) branch, and there
is no ``graceful exit''. The egg can only convert
($-$) to (+). Clearly, multiple eggs cannot
change this conclusion.

To prove the above statements, one notices that for the first
case the integral formula
applied for the $(-)$ bounce between the hit point $(t_0)$ and the
terminus of the egg $(t_1)$ rules this option out.
In this region the flow is to the right and $h \ge 0$.
Therefore, the integral is equal to the area between
the segment and the $\phi$-axis and hence strictly positive
$\int_{\phi(t_0)}^{\phi(t_1)} h d\phi = {\cal A} >0$
Substituting the corresponding parameters in the integral formula
we arrive at the sought
contradiction:
\begin{equation}
0<2{\cal A}=-(\sqrt{3} -1) h(t_1) - h(t_0) \le 0
\label{contra1}
\end{equation}
Therefore, the ($-$) bounce emerging from the upper side of the
egg must rehit it, as we claimed.
A similar contradiction shows the (+) trajectory cannot hit below
the egg. Furthermore, there may be several complications regarding
the pathological
trajectories hitting the singular egg points, corresponding to
inflections of the
potential or local maxima at $\Lambda=0$. Although these are
not covered by the above arguments, they do not lead to
graceful exit either \cite{kmo}.

As we have mentioned before, this no-go theorem can be generalized
to the case when stringy fluid sources are present. In this case, the phase
space of the model should be extended to three
dimensions, the third coordinate being the energy density of the fluid
$\rho$. The associated equations of motion are given by the following
generalization of (\ref{bv1}-\ref{bv2}) \cite{bv, kmo}:
\begin{eqnarray}
\dot \phi &=& (3 h \pm \sqrt{3 h^2+2 \Lambda(\phi)
+ \rho \exp{(2\phi)}})/2 \nonumber \\
\noalign{\medskip}
\dot h   &=& \pm h \sqrt{3 h^2+2 \Lambda(\phi) + \rho \exp{(2\phi)}}
-\Lambda'(\phi)/2 + \frac{\gamma}{2} \rho \exp{(2\phi)} \nonumber \\
\noalign{\medskip}
\dot \rho &=& - 3(1 + \gamma) h \rho
\label{source}
\end{eqnarray}
Here $\gamma = p/\rho \in (-1/3, 1/3)$ is a constant representing the
fluid equation of state. The physical restriction $\rho \ge 0$
is consistent with the equations of motion, as the $\rho$ flow terminates at
the $\rho=0$ plane, which is like a potential barrier.
The trajectories completely confined in this plane are governed by our
previous theorem, so there is no graceful exit for them. Now we look at
the fully three-dimensional trajectories. The egg function is given by
$e=\sqrt{3 h^2+2 \Lambda(\phi) + \rho \exp{(2\phi)}}$ and
the modified integral formula when sources
are present is
\begin{eqnarray}
\pm(e(t_1)-e(t_0))+h(t_1)-h(t_0)=2 \int_{\phi(t_0)}^{\phi(t_1)} h d\phi
+ \frac{1+\gamma}{2} \int_{t_0}^{t_1} \rho e^{2\phi} dt
\label{diffintfl}
\end{eqnarray}
This equation differs from (\ref{diffint}) only in the presence of the last
term, which is a nonnegative quantity for all trajectories. The other
relevant characteristics of the phase space are also easily generalized.
Without delving into the details, we note
that qualitatively the picture remains the same: the egg is now a
two-dimensional compact surface cut by the plane $\rho=0$.
Trajectories flow around the egg along helical
paths, which if projected onto the $\rho=0$ plane turn clockwise. We then
need to consider trajectories crossing the cylindrical surface enclosing the
egg, obtained by translating the curve representing the boundary of the egg
in the $h=0$ plane vertically upwards. We can see that the first integral on
the LHS, representing the area enclosed by
the projection of the trajectory onto the $\rho=0$ plane, remains positive
for all such trajectories. The second integral is always
nonnegative, as we mentioned above. Thus all the arguments for the
sourceless case extend to this case, again preventing favorable branch
changes from occurring. As no new pathologies appear, we conclude that the
no-go theorem must hold for this case too.

We note in passing that dilaton-dependent
corrections to couplings, coming from the
higher genus terms \cite{dp,dv}, also fail
to induce the desired graceful exit \cite{kmo}.
In that case, the failure is caused solely by the flows around the eggs which
preclude even the possibility for the needed hits on the egg.
Thus the only option to consistently incorporate
Pre-Big-Bang cosmology which remains open is to resort to higher derivative
terms in the $\alpha'$ expansion \cite{ck}. However this approach must consider
systematically all the terms in the $\alpha'$ expansion, and thus should be
implemented via the conformal field theory approach. At this moment,
it appears that this goal is still beyond our means.

\vskip 0.5truecm
\noindent {\bf Acknowledgements}
\vskip 0.5truecm
The author wishes to thank R. Madden and K.A.
Olive for a fruitful collaboration and many helpful discussions.
This report was supported in part by NSERC of Canada and in
part by an NSERC postdoctoral
fellowship.

\end{document}